\documentclass[sigconf]{acmart}

\usepackage{graphicx}

\usepackage{xcolor}

\usepackage{algorithm}
\usepackage[noend]{algpseudocode}

\algrenewcommand\alglinenumber[1]{#1}

\usepackage{multirow}

\usepackage{enumitem}
\setlist[itemize]{leftmargin=*, topsep=.0em, itemsep=0pt, parsep=0pt, partopsep=0pt}
\setlist[enumerate]{leftmargin=*, topsep=.0em, itemsep=0pt, parsep=0pt, partopsep=0pt}

\usepackage{atbegshi}
\AtBeginDocument{\AtBeginShipoutNext{\AtBeginShipoutDiscard}}
\addtocounter{page}{-1}

\usepackage{amsthm}
\newtheoremstyle{def_style}
  {.0em}          
  {.0em}          
  {}          
  {}          
  {\bfseries} 
  {.}         
  {.0em}      
  {}          

\theoremstyle{def_style}

\theoremstyle{def_style}
\newtheorem{prob}{Problem}

\copyrightyear{2023}
\acmYear{2023}
\setcopyright{acmlicensed}
\acmConference[CIKM '23]{Proceedings of the 32nd ACM International Conference on Information and Knowledge Management}{October 21--25, 2023}{Birmingham, United Kingdom}
\acmBooktitle{Proceedings of the 32nd ACM International Conference on Information and Knowledge Management (CIKM '23), October 21--25, 2023, Birmingham, United Kingdom}
\acmPrice{15.00}
\acmDOI{10.1145/3583780.3614655}
\acmISBN{979-8-4007-0124-5/23/10}

\begin{document}

\title{An Efficient Content-based Time Series Retrieval System}
\author{Chin-Chia Michael Yeh, Huiyuan Chen, Xin Dai, Yan Zheng, Junpeng Wang, Vivian Lai, Yujie Fan, Audrey Der$^\dag$, Zhongfang Zhuang, Liang Wang, Wei Zhang, and Jeff M. Phillips$^\ddag$}
\affiliation{%
    \institution{Visa Research, University of California, Riverside$^\dag$, University of Utah$^\ddag$}
}


\renewcommand{\shortauthors}{Chin-Chia Michael Yeh et al.} 

\begin{abstract}
A \textit{Content-based Time Series Retrieval} (CTSR) system is an information retrieval system for users to interact with time series emerged from multiple domains, such as finance, healthcare, and manufacturing.
For example, users seeking to learn more about the source of a time series can submit the time series as a query to the CTSR system and retrieve a list of relevant time series with associated metadata.
By analyzing the retrieved metadata, users can gather more information about the source of the time series.
Because the CTSR system is required to work with time series data from diverse domains, it needs a high-capacity model to effectively measure the similarity between different time series. 
On top of that, the model within the CTSR system has to compute the similarity scores in an efficient manner as the users interact with the system in real-time.
In this paper, we propose an effective and efficient CTSR model that outperforms alternative models, while still providing reasonable inference runtimes.
To demonstrate the capability of the proposed method in solving business problems, we compare it against alternative models using our in-house transaction data.
Our findings reveal that the proposed model is the most suitable solution compared to others for our transaction data problem.
\end{abstract}

.


\maketitle
\thispagestyle{empty}
\section{Introduction}
Time series is a common data type analyzed for a variety of applications~\cite{dau2019ucr}.
For example, time series from different sensors on manufacturing machines are examined by engineers for identifying ways to improve factories' efficiency, various biometric time series are studied by doctors for medical research, and multiple streams of time series from operating payment network are monitored for unusual activities.
As a large volume of time series data are becoming available from various sources, it is essential to develop an effective system to help users browse time series databases. 
In this paper, we refer to such a system as a \textit{Content-based Time Series Retrieval} (CTSR) system. 
This system is designed to retrieve relevant time series from a database when given a query time series.

\begin{figure}[ht]
\centerline{
\includegraphics[width=0.95\linewidth]{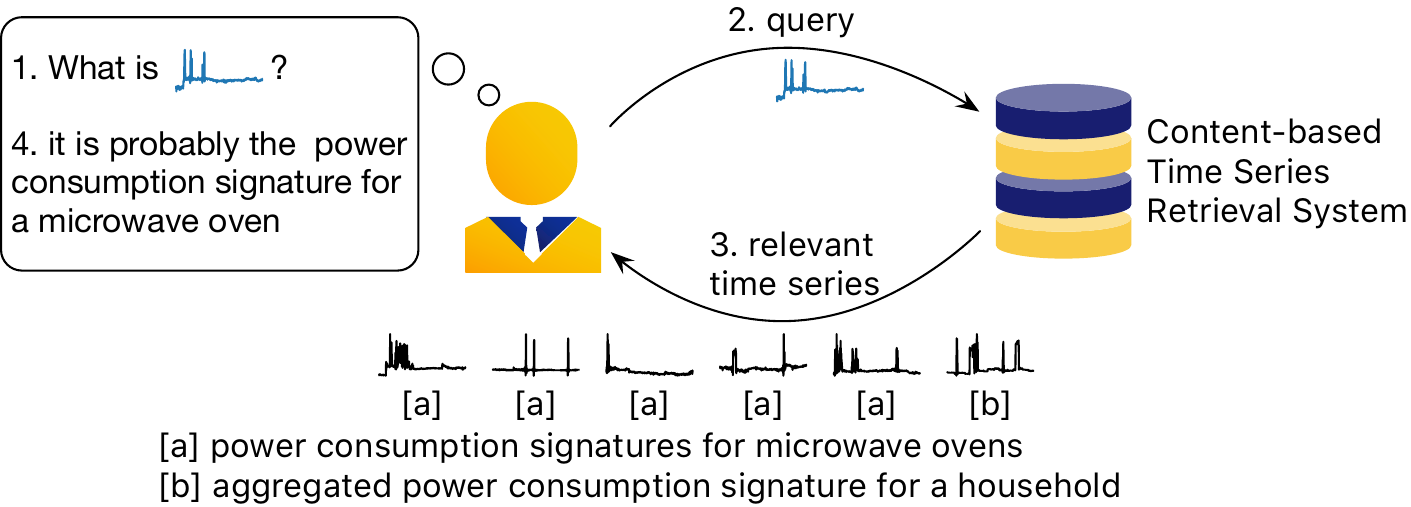}
}
\caption{
The use case for a \textit{Content-based Time Series Retrieval} (CTSR) system where the database consists of time series from multiple domains.
}
\label{fig:motivation_1}
\end{figure}

To understand what a CTSR system is and how it can help users, let us consider the use case presented in Figure~\ref{fig:motivation_1}. 
Suppose a user comes across a time series without any associated meta data.
The time series could be a power consumption time series or data records from other sensors.
The user wants to identify the possible source of the time series and recover the missing information. 
To solve this problem, the user queries the CTSR system\footnote{The query time series may or may not exist in the database of the CTSR system.} with the time series, and the system returns a ranked list of similar time series with associated meta data.
In this example, five out of the top six returned time series are power consumption signatures for microwave ovens.
Consequently, the user is able to infer that the unknown time series is most likely a microwave oven's power consumption signature.
Hence, the CTSR system helps the user in recovering the missing information about time series.

We have two primary design goals when building our CTSR system:
1) to effectively capture various concepts in time series from different domains, and
2) to be efficient during inference, given the real-time interactions of users with the system.
To identify the most suitable distance or similarity function for our system, we conducted a benchmark experiment using two distance functions and five neural network models (see Section~\ref{sec:method}).
From the seven tested methods, the \textit{Residual Network $2D$} (RN2D) outperforms the other methods with statistical significance.
However, the RN2D method failed to meet the second design goal. 
Its average query time was 32 seconds, while most of the other rival methods' query times were below 0.04 seconds (refer to Table~\ref{tab:ucr_result}). 
In other words, while the RN2D method proved to be the best model in terms of accuracy, it may not be ideal for building our system if we aim to guarantee a reasonable response latency.

\begin{figure}[ht]
\centerline{
\includegraphics[width=0.95\linewidth]{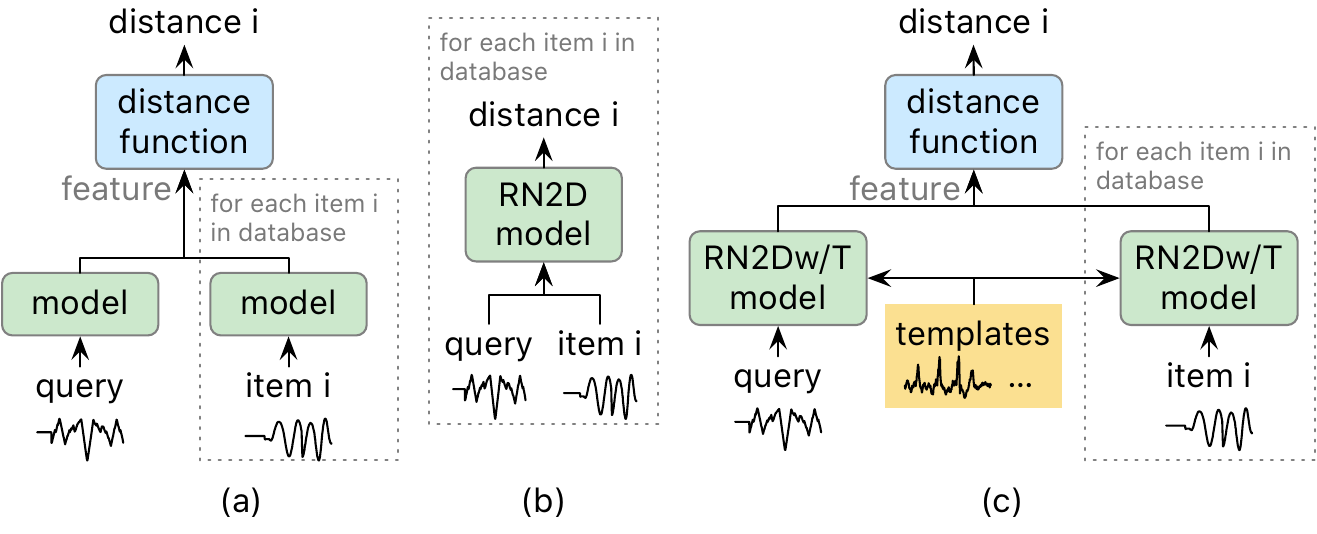}
}
\caption{
The design of the neural network can greatly impact the inference time.
}
\label{fig:style}
\end{figure}

The main reason for the drastic difference in inference time is the difference in the role of the neural network. 
In faster methods (i.e., the methods with query time less than 0.04 seconds), the neural network serves purely as a \textit{feature extractor}, and the distance is computed using Euclidean distance (see Figure~\ref{fig:style}.a).
Therefore, we only need to project each time series in the database to Euclidean space once using the neural network before inference. 
During query time, the neural network only needs to project the query time series to the same Euclidean space, and the distance computation can be efficiently performed in this space.
On the other hand, the RN2D model serves as both the \textit{feature extractor} and \textit{distance function} as shown in Figure~\ref{fig:style}.b. 
Thus, the RN2D model is invoked each time the distance is computed. 
In cases where there are $n$ time series in the database, the faster method only requires one invocation of the neural network for the query. 
In contrast, the RN2D model invokes $n$ times, drastically increasing the runtime.

To leverage the advantages of the best available model (RN2D) for our CTSR system, we propose a novel model architecture based on the RN2D model with improved efficiency, called \textit{Residual Network $2D$ with Template Learning} (RN2Dw/T). 
As illustrated in Figure~\ref{fig:style}.c, we incorporate a template (landmark) learning mechanism into the input of the RN2D model (Figure~\ref{fig:style}.b) and modify the model to output feature vectors instead of distance values. 
Unlike the RN2D method, the RN2Dw/T model functions solely as a \textit{feature extractor}. 
Our proposed RN2Dw/T achieves comparable effectiveness to the RN2D method while maintaining an average query time of less than 0.04 seconds (see Table~\ref{tab:ucr_result}).
Our results provide strong evidence that an effective and efficient CTSR system with the proposed RN2Dw/T model can be a valuable tool for businesses in various industries.


\section{Background}
\label{sec:background}
In this section, we will begin by providing the problem statement, and then we will review relevant works from the literature.

\vspace{-0.5em}

\subsection{Problem Statement}
The CTSR problem is formulated as follows:

\begin{prob}
\label{prob:ctsr}
Given a set of time series~$\mathcal{X}=[\mathbf{x}_1,\cdots,\mathbf{x}_n]$ and any query time series~$\mathbf{q}$, we want to obtain a relevance score function~$f(\cdot,\cdot)$, which satisfies the property that $f(\mathbf{x}_i, \mathbf{q}) > f(\mathbf{x}_j, \mathbf{q})$ if $\mathbf{x}_i$ is more relevant to $\mathbf{q}$ than $\mathbf{x}_j$.
\end{prob}

\noindent The scoring function can be either a predefined similarity function or a trainable function that is optimized using the metadata associated with each time series in $\mathcal{X}$.

\subsection{Related Work}
There are two major ways to formulate the time series retrieval problem~\cite{muhammad2010multi,esling2012time,rakthanmanon2012searching,song2018deep,zhu2020deep}. 
The first is known as the time series similarity search problem, which focuses on finding the top $k$ time series that are most similar to a given query based on a fixed distance function~\cite{muhammad2010multi,rakthanmanon2012searching,camerra2010isax,palpanas2020evolution,zhu2020deep}.  
For efficiency, techniques such as lower bounding~\cite{rakthanmanon2012searching}, early abandoning~\cite{rakthanmanon2012searching}, and/or indexing~\cite{palpanas2020evolution} are commonly used. 
This type of research differs from our goal (i.e., Problem~\ref{prob:ctsr}).

The second type of problem formulation is more aligned with our problem statement, where the objective is to develop a model or scoring function to aid users in retrieving relevant time series from a database based on the query time series submitted.
We searched for relevant literature and found only one paper~\cite{song2018deep}.
Their proposed model was intended for multivariate time series analysis. 
If we were to adapt it for our univariate time series problem, the resulting model would essentially be a standard long short-term memory network~\cite{hochreiter1997long}. 
Consequently, we expanded our literature review further to identify effective distance measures and models for time series data, regardless of their specific application.

The Euclidean distance and dynamic time warping distance are popular tools for analyzing time series data. 
They are widely used in various tasks such as similarity search, classification, and anomaly detection~\cite{esling2012time, rakthanmanon2012searching, yeh2016matrix, dau2019ucr, lu2022matrix}. 
We can readily apply both distance functions to our problem, and therefore, we include them in our experiments.
Another family of methods that can be used in our problem is neural networks. 
For example, long short-term memory networks~\cite{hochreiter1997long, karim2017lstm, karim2019multivariate, lim2021time, zhou2021informer}, gated recurrent unit networks~\cite{cho2014properties, lim2021time, zhou2021informer}, transformers~\cite{vaswani2017attention, li2019enhancing, zhou2021informer, lim2021time, chen2022denoising}, and convolutional neural networks~\cite{wang2017time, ismail2019deep, ren2019time} have shown effectiveness in tasks such as time series classification, forecasting, and anomaly detection. 
Thus, we also include these neural network models in our experiments.

\section{Method}
\label{sec:method}
In this section, we begin by presenting six common baseline methods. 
Following that, we introduce the Residual Network $2D$ (RN2D) method and contrast its benefits with the other baseline methods. 
Once we have introduced the RN2D method, we will present the proposed Residual Network $2D$ with Template learning (RN2Dw/T) method, which solves the efficiency issue associated with the design of RN2D.
The six common baseline methods are:

\begin{itemize}
\item \textbf{Euclidean Distance (ED)}:
We compute the Euclidean distance between the query time series and the time series in the collection. 
Then, we sort the collection based on the distances. 
This is the simplest approach for solving the CTSR problem.
\item \textbf{Dynamic Time Warping (DTW)}: 
This method is similar to the ED baseline, but uses the DTW distance instead. 
The DTW distance is considered as a simple yet effective baseline for time series classification problems~\cite{rakthanmanon2012searching,bagnall2017great,dau2019ucr}. 
Therefore, it is crucial to include this method in our CTSR benchmark.
\item \textbf{Long Short-Term Memory network (LSTM)}: 
The LSTM is one of the most popular Recurrent Neural Networks (RNNs) used for modeling sequential data~\cite{hochreiter1997long,lim2021time,zhou2021informer}. 
In this work, we optimize LSTM models using the Siamese network architecture~\cite{chicco2021siamese} (see Figure~\ref{fig:style}.a). 
\item \textbf{Gated Recurrent Unit network (GRU)}: 
The GRU is another popular RNN architecture widely used for modeling sequential data~\cite{cho2014properties,lim2021time,zhou2021informer}. 
To optimize the GRU model, we applied a similar approach as for the LSTM model, wherein we replaced the LSTM cells in the RNN architecture with GRU cells.
\item \textbf{Transformer (TF)}: 
The transformer is an alternative to the RNNs for sequence modeling~\cite{vaswani2017attention,li2019enhancing,zhou2021informer,lim2021time,chen2022denoising}.
To learn the hidden representation for the input time series, we utilized the transformer encoder proposed in~\cite{vaswani2017attention}. 
We replaced the RNNs used in the previous two methods (i.e., LSTM and GRU) with transformer encoders, resulting in a transformer-based Siamese network architecture instead of an RNN-based one.
\item \textbf{Residual Network 1D (RN1D)}: 
The RN1D is a time series classification model inspired by the success of residual networks in computer vision~\cite{he2016deep,wang2017time}. 
It employs $1D$ convolutional layers instead of $2D$ convolutional layers~\cite{he2016deep,wang2017time} and is first proposed in~\cite{wang2017time}. 
Extensive evaluations conducted by~\cite{ismail2019deep} have demonstrated that the RN1D design is among the strongest models for time series classification. 
We once again optimize this model in a Siamese network fashion~\cite{chicco2021siamese} (see Figure~\ref{fig:style}.a). 
\end{itemize}

Both the ED and DTW methods require no training phase as they have no parameters to optimize. 
The DTW method is the more effective method of the two for time series data, because it considers all alignments between the input time series~\cite{rakthanmanon2012searching}.
The computation of DTW distance can be abstracted into a two-stage process.
In the first stage, a pair-wise distance matrix $D \in \mathbb{R}^{w \times h}$ is computed from the input time series $\mathbf{a}=[a_1, \cdots, a_w]$ (where $w$ is the length of $\mathbf{a}$) and $\mathbf{b}=[b_1, \cdots, b_h]$ (where $h$ is the length of $\mathbf{b}$) as $D[i, j]=|a_i-b_j|$. 
In the second stage, a fixed recursion function is applied to $D$, i.e., $D[i,j] \gets D[i,j] + \textsc{min}(D[i-1, j], D[i, j-1], D[i-1, j-1])$, for each element in $D$. 
Consequently, the DTW method can be viewed as running a predefined function on the pair-wise distance matrix between the input time series.

The remaining four baseline methods use the Siamese network distance learning framework (refer to Figure~\ref{fig:style}.a) and employ high-capacity\footnote{We use the term \textit{capacity} to describe the \textit{expressiveness}~\cite{lu2017expressive} of a model.} neural network models, i.e., LSTM, GRU, TF, and RN1D, to learn the hidden representation of the input time series. 
These representations are used to compute the distance between two time series, and the models are learned using the optimization procedure detailed in Section~\ref{sec:optimize}. 
Once the model is optimized, the hidden representations of each time series in the database are extracted before deployment. 
When a user submits a query, we only need to apply the model to the query time series to extract its hidden representation because the hidden representations of each time series in the database are already extracted before query time. 
Then, the distances between the query and each item in the database are computed using the Euclidean distance.

\subsection{Residual Network 2D}
\begin{figure}[ht]
\centerline{
\includegraphics[width=0.95\linewidth]{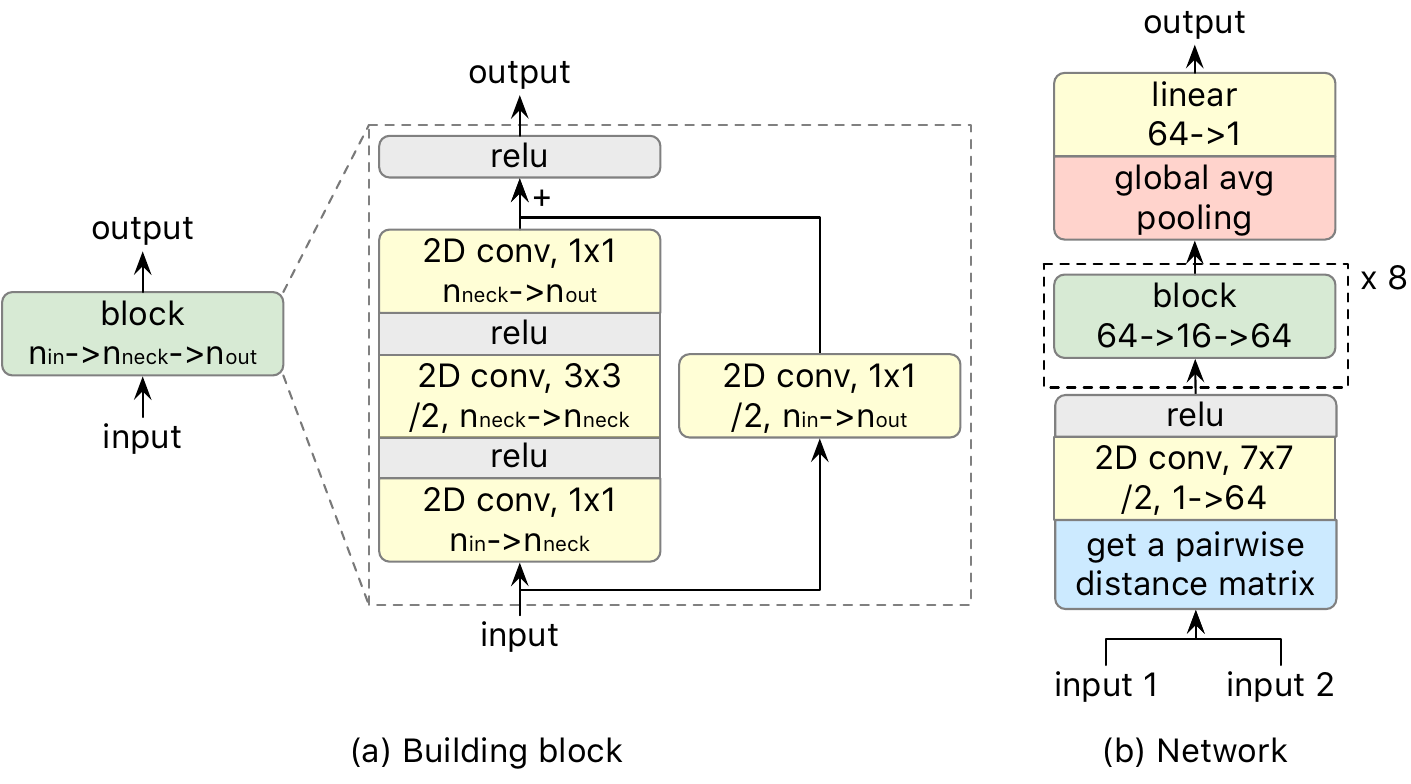}
}
\caption{
The building block and network designs for the Residual Network $2D$ (RN2D) model are shown in Figure~\ref{fig:resnet_2d}.a and Figure~\ref{fig:resnet_2d}.b, respectively.
}
\label{fig:resnet_2d}
\end{figure}

The Residual Network 2D (RN2D) method (Figure~\ref{fig:resnet_2d}) combines the best of both worlds (i.e., DTW and neural network).
The RN2D method takes advantage of the rich alignment information from the pair-wise distance matrix, similar to the DTW method. 
However, instead of using a fixed function, the RN2D method uses a high-capacity neural network as the function, making use of an expressive model like the four neural network baselines. 
Figure~\ref{fig:resnet_2d}.a shows the building block, while Figure~\ref{fig:resnet_2d}.b shows the overall model.
Please see~\cite{supplementary} for a detailed description of each module in Figure~\ref{fig:resnet_2d}.

As shown in Figure~\ref{fig:style}.b, we cannot extract the hidden representations of each time series in the database before deployment when using RN2D, unlike the methods using the Siamese network framework. 
If there are $n$ time series in the database, we need to run RN2D $n$ times during query time to compute the distance between the query time series and each time series in the database. 
In contrast, the methods using the Siamese network framework only need to run the model once during query time, making the RN2D method an order of magnitude slower\footnote{The claim is based on our observation of the experimental results presented in Table~\ref{tab:ucr_result}.}. 
To address the efficiency issue of RN2D, we propose the Residual Network $2D$ with Template Learning method, which we will introduce in the next section.

\subsection{Residual Network 2D with Template Learning}
\begin{figure}[ht]
\centerline{
\includegraphics[width=0.85\linewidth]{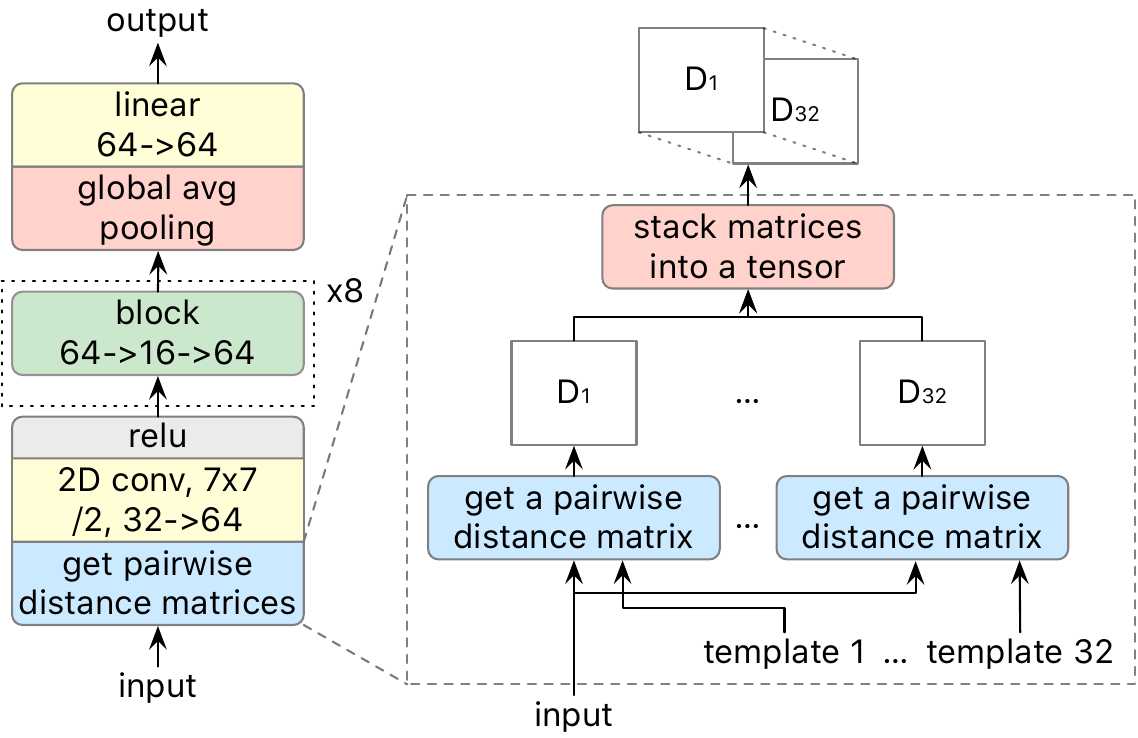}
}
\caption{
The proposed Residual Network $2D$ with Templates Learning (RN2Dw/T) model.
}
\label{fig:resnet_t}
\end{figure}

We propose the Residual Network $2D$ with Template Learning (RN2Dw/T) to address the efficiency issue of RN2D. 
This method is designed to be as effective as the RN2D method while being an order of magnitude faster.
Figure~\ref{fig:resnet_t} illustrates the design of the RN2Dw/T model.

The proposed RN2Dw/T model begins by computing a set of pairwise distance matrices between the input time series and a set of templates. 
These distance matrices are computed as follows: 
Given an input time series $\mathbf{a}=[a_1, \cdots, a_w]$ and the $k$th template $\mathbf{t}_k=[t_{k,1}, \cdots, t_{k,w}]$, the $k$th distance matrix $D_k \in \mathbb{R}^{w \times h}$ is computed with $D_k[i, j]=|a_i-t_{k,j}|$. 
We compute the distance matrix for each of the 32 templates, resulting in 32 $w \times h$ matrices. 
The 32 templates are learned during the training phase as part of model parameters and are reference time series that help the model project the input time series to Euclidean space using the RN2D design.
Next, we stack the 32 $w \times h$ matrices together to create a tensor, $D \in \mathbb{R}^{w \times h \times 32}$. 
This tensor is the output of the pairwise distance matrix computation step.

Similar to the RN2D model (as shown in Figure~\ref{fig:resnet_2d}.a), we employ a $7 \times 7$ convolutional layer with a step size of two to project the tensor $D$ onto $\mathbb{R}^{w/2 \times h/2 \times 64}$ space.
After applying a ReLU layer, the resulting intermediate representation passes through eight building blocks with the $64{\to}16{\to}64$ configuration.
Next, we apply a global average pooling layer to reduce the spatial dimensions, resulting in a size-64 vector. 
Finally, we use a linear layer to further process this vector into a size-64 vector, which serves as the final representation of the input time series.

For RN2Dw/T model, the feature of the input time series is computed based on the pairwise distance matrices between the input time series and the learned templates.
This allows us to use the trained RN2Dw/T model to extract the feature vectors for all the items in the database before deployment, and we only need to do this feature extraction step once.
After deployment, when a query time series is obtained, we only need to extract its feature vector, after which we can efficiently compute the Euclidean distance between the query feature vector and the pre-computed feature vectors from the database.
However, if the RN2D model is used in the system, we would need to reprocess all items in the database whenever we receive a new query, which would considerably increase the query time.

The proposed RN2Dw/T method bears some resemblance to the Nystr\"om approximation method for kernel learning ~\cite{williams2000using,drineas2005nystrom}, in that both involve using a subset of samples to replace the training dataset. 
This similarity highlights how RN2Dw/T can achieve similar performance to RN2D, just as the Nystr\"om method can approximate the exact kernel.
The RN2Dw/T model has a similar capacity to the RN2D model, but it also enables a more efficient query mechanism, which is crucial when designing a real-world CTSR system.

\subsection{Optimization}
\label{sec:optimize}
\sloppy
The loss function used in our study is the BPR loss~\cite{rendle2009bpr}, which is appropriate for the CTSR problem since it is a \textit{learning-to-rank} problem. 
Given a batch of training data $\mathcal{B}=[\mathcal{T}_0, \cdots, \mathcal{T}_m]$, the loss function is defined as:$\sum_{(\mathbf{t}_i, \mathbf{t}_{i,+}, \mathbf{t}_{i,-})\in \mathcal{B}} -\log \sigma(f_{\theta}(\mathbf{t}_i, \mathbf{t}_{i,+}) - f_{\theta}(\mathbf{t}_i, \mathbf{t}_{i,-})),$
where $m$ is the batch size, $\sigma(\cdot)$ is the sigmoid function, and $f_{\theta}(\cdot,\cdot)$ is the model. 
Each sample in the batch is a tuple~$\mathcal{T}_i$ that contains the query (or anchor) time series~$\mathbf{t}_i$, the positive time series~$\mathbf{t}_{i,+}$, and the negative time series~$\mathbf{t}_{i,-}$.
We used the AdamW optimizer~\cite{loshchilov2018decoupled} to train our models.

\section{Experiment}
\label{sec:experiment}
In this section, we present the results of our experiments on a CTSR benchmark dataset created from the UCR Archive~\cite{dau2019ucr} and an in-house transaction dataset based on a real business problem (see Figure~\ref{fig:motivation_0}).
Neural network-based methods were implemented using PyTorch~\cite{paszke2019pytorch}, and the model with the best average NDCG@10 score on the validation data was selected for testing.
We used SciPy~\cite{virtanen2020scipy} to compute ED and Tslearn~\cite{tavenard2020tslearn} to compute DTW.
Further details including source code and hyper-parameter settings can be found in the project website~\cite{supplementary}.

\subsection{UCR Archive}
\label{sec:ucr_archive}
We created the CTSR benchmark dataset from the UCR Archive~\cite{dau2019ucr}, which is a collection of 128 time series classification datasets from various domains such as motion, power demand, and traffic. 
The UCR Archive is widely used for benchmarking time series classification algorithms~\cite{bagnall2017great,wang2017time,ismail2019deep}. 
We convert the UCR Archive to a CTSR benchmark dataset using the steps listed on the project website~\cite{supplementary}.
The resulting dataset consists of 136,377 training time series, 17,005 test queries, and 17,005 validation queries.

To measure the performance of different retrieval methods, we computed common information retrieval metrics, including precision at $k$ (Prec@$k$), average precision at $k$ (AP@$k$), and normalized discounted cumulative gain at $k$ (NDCG@$k$), for each query.
The performance measurements at $k=10$ for each of the 17,005 test queries are averaged and presented in Table~\ref{tab:ucr_result}. 
When comparing different methods, we also conducted two-sample t-tests with $\alpha=0.05$ to test for statistical significance.
Please refer to~\cite{supplementary} for the complete results of the significance tests.
The reported query time is the average time taken to compute relevant scores between a query and the 136,377 time series in the training dataset. 
We computed the average query time by using 1,000 different time series from the test data as the query. 
Table~\ref{tab:ucr_result} allows for easy comparison of different methods based on different performance measures.

\begin{table}[ht]
\centering
\caption{
The proposed RN2Dw/T method is both effective and efficient.
The query times are measured in seconds.
}
\label{tab:ucr_result}
\footnotesize
\begin{tabular}{l||ccc|c}
Method & PREC@10 & AP@10 & NDCG@10 & Query Time \\ \hline \hline
ED & 0.7316 & 0.7655 & 0.7499 & 0.0353 \\
DTW & 0.8562 & 0.8792 & 0.8693 & 36.1224 \\ \hline
LSTM & 0.9205 & 0.9277 & 0.9260 & 0.0169 \\
GRU & 0.9221 & 0.9285 & 0.9273 & 0.0084 \\
TF & 0.9146 & 0.9212 & 0.9199 & 0.0029 \\
RN1D & 0.9086 & 0.9164 & 0.9146 & 0.0366 \\ \hline
RN2D & 0.9266 & 0.9342 & 0.9325 & 32.0752 \\
RN2Dw/T & \textbf{0.9286} & \textbf{0.9343} & \textbf{0.9336} & 0.0181
\end{tabular}
\end{table}

Firstly, we focus our discussion on the three performance measurements: PREC@10, AP@10, and NDCG@10. 
When comparing the performance of the two non-neural network baselines (ED and DTW), we observed that DTW significantly outperforms ED in all three performance measurements.
This suggests that using alignment information helps with the CTSR problem, and similar conclusions have been drawn for the time series classification problem~\cite{bagnall2017great}.

When considering the first four neural network baselines (i.e., LSTM, GRU, TF, and RN1D), all of them significantly outperform the DTW method. 
This demonstrates that using a high-capacity model helps with the CTSR problem. 
One possible reason for this is that the CTSR dataset consists of time series from many different domains~\cite{dau2019ucr}, and higher capacity models are required for learning diverse patterns within the data.
Among the four methods, GRU and LSTM outperform the other two methods significantly in all three performance measurements; GRU performs slightly better compared to LSTM, but the difference is not statistically significant.

The RN2D method, a high-capacity model utilizing alignment information, significantly outperforms all other methods according to the t-test results. 
When comparing the RN2Dw/T method with the RN2D method, the RN2Dw/T method achieves higher performance in all three performance measurements, although the difference is not significant. 
Thus, both the proposed RN2Dw/T method and the RN2D method can be considered the best performing methods for the CTSR dataset in terms of the three performance measurements.

When considering the query time, the eight tested methods can be grouped into two categories: slower methods (i.e., DTW and RN2D) with a query time of over 30 seconds, and faster methods (i.e., ED, LSTM, GRU, TF, RN1D, and RN2Dw/T) where each query takes less than 0.1 seconds. 
The main difference between the faster and slower groups is that all fast methods compute the relevance score in Euclidean space, while the slower methods compute the scores in non-Euclidean spaces. 
Overall, the proposed RN2Dw/T method is the best method as it is effective in retrieving relevant time series and efficient in terms of query time.

\begin{figure}[ht]
\centerline{
\includegraphics[width=0.85\linewidth]{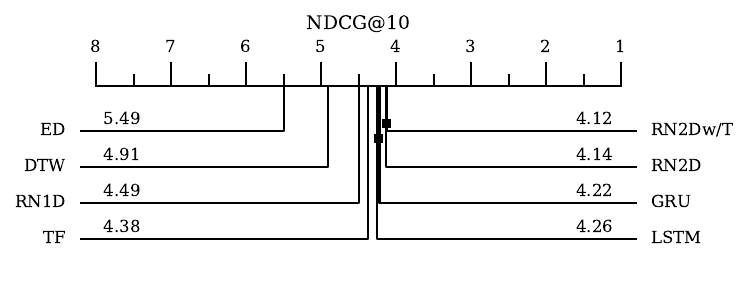}
}
\caption{
We also construct the CD diagram for PREC@10 and PR@10, the conclusion remains the same.
}
\label{fig:ucr_cd_diag}
\end{figure}

Following many prior works in time series classification~\cite{bagnall2017great,ismail2019deep}, we constructed a critical difference (CD) diagram to compare the performance of different methods. 
The CD diagram (see Figure~\ref{fig:ucr_cd_diag}) shows the average rank of each method based on a performance measurement and indicates whether two methods exhibit a significant difference in performance based on the Wilcoxon signed-rank test ($\alpha=0.05$). 
The conclusion is consistent with the findings presented in Table~\ref{tab:ucr_result}.

\begin{figure}[ht]
\centerline{
\includegraphics[width=0.8\linewidth]{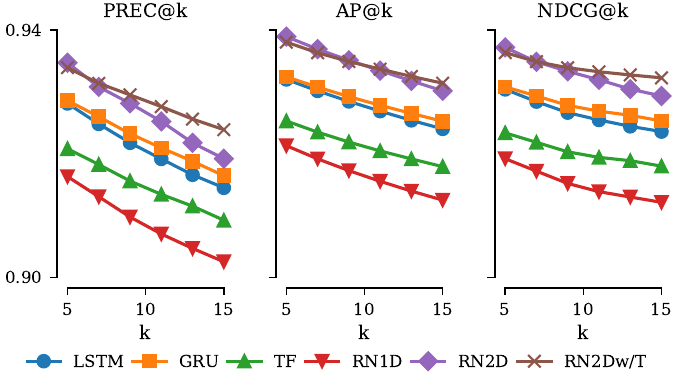}
}
\caption{
The proposed method outperforms the others with different settings of $k$ for each performance measure.
}
\label{fig:ucr_k_plot}
\end{figure}

In Figure~\ref{fig:ucr_k_plot}, we present the results of the performance measurements using various values of $k$ ranging from 5 to 15. 
This is done to ensure that the conclusions drawn from Table~\ref{tab:ucr_result} and Figure~\ref{fig:ucr_cd_diag} are not limited to our particular choice of $k$.
To improve readability, we have omitted ED and DTW from the figures since their performance is much worse than the other methods.

As shown in Figure~\ref{fig:ucr_k_plot}, the proposed RN2Dw/T method achieves the best performance across different values of $k$ for all three performance measurements. 
The order of the remaining methods from best to worst is: RN2Dw/T, RN2D, GRU, LSTM, TF, and RN1D. 
These results are consistent with the findings presented in Table~\ref{tab:ucr_result} and Figure~\ref{fig:ucr_cd_diag}.

\begin{figure}[ht]
\centerline{
\includegraphics[width=0.8\linewidth]{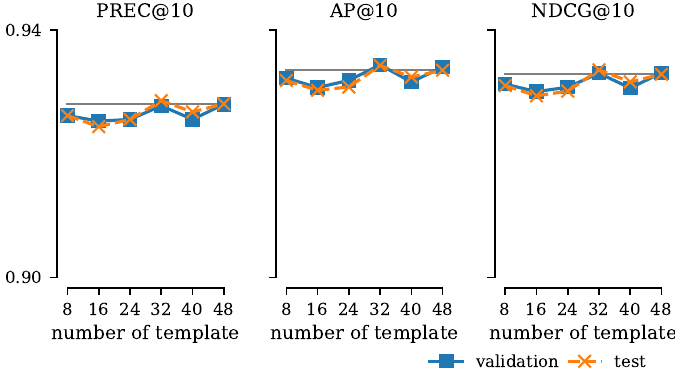}
}
\caption{
The proposed RN2Dw/T method is not sensitive to the number of template settings. 
The gray horizontal line marks the test performance of the RN2D method.
}
\label{fig:ucr_n_temp}
\end{figure}

The proposed RN2Dw/T method introduces an additional hyper-parameter: the number of templates used to compute the pairwise distance matrices (see Figure~\ref{fig:resnet_t}). 
We experimented with various settings, including 8, 16, 24, 32, 40, and 48, and the results are presented in Figure~\ref{fig:ucr_n_temp}. 
The performance of RN2Dw/T is similar to that of the RN2D method, regardless of the hyper-parameter setting.
One potential reason for the lack of sensitivity of RN2Dw/T to this hyper-parameter setting is that the performance of the learned representation is not influenced by the specific shape of templates. 
A similar phenomenon can be observed in various shapelet-based methods~\cite{renard2016east,dempster2020rocket,yeh2018towards}.
We chose to set the number of templates to 32 based on its performance on the validation set.

\begin{figure}[ht]
\centerline{
\includegraphics[width=0.85\linewidth]{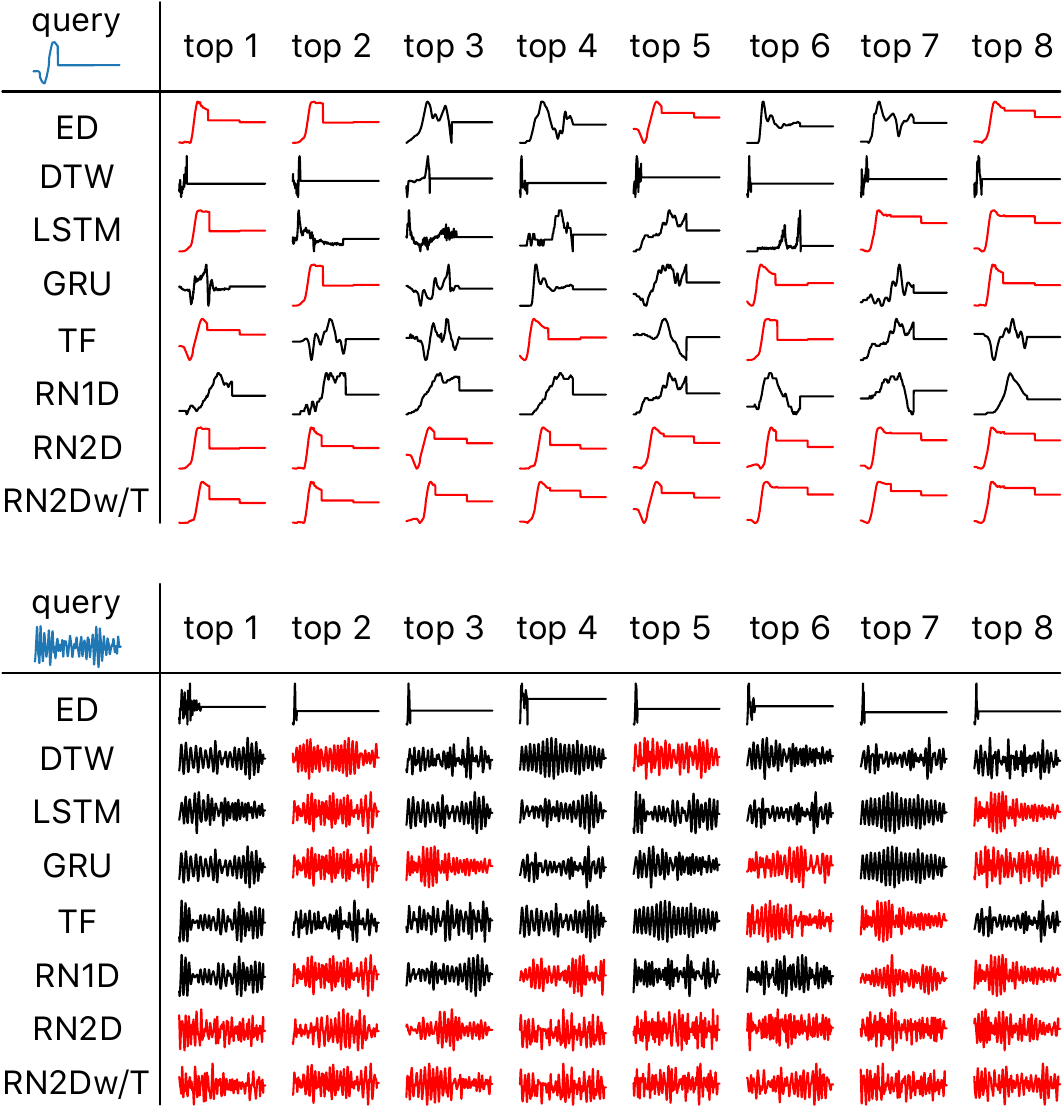}
}
\caption{
The top eight items of a given query time series were retrieved by different methods.
The retrieved time series is colored \textcolor{red}{red} if it is relevant and black if it is irrelevant.
}
\label{fig:ucr_query}
\end{figure}

Next, in Figure~\ref{fig:ucr_query}, we present the top eight retrieved time series using different methods. 
We selected two queries with different levels of complexity from the test dataset. 
The simpler query consists of a single cycle of a pattern, while the complex query contains periodic signals. 
Periodic signals in complex queries typically require shift-invariant distance measures~\cite{paparrizos2015k} to retrieve relevant items correctly. 
This figure demonstrates that the CTSR problem is challenging, as even irrelevant time series are visually similar to the query.

Examining the retrieved time series for different methods, we make the following observations: 
The ED method struggles with the more complex query because it cannot align the query to relevant time series. 
The DTW method outperforms the ED method on the complex query, but its alignment freedom hurts its performance on the simple query. 
The four neural network baselines (i.e., LSTM, GRU, TF, and RN1D) perform better than both ED and DTW methods when considering both queries. 
However, none of these baselines outperforms both the RN2Dw/T and RN2D methods, which reliably retrieve relevant items.

\subsection{Transaction Time Series}
\begin{figure}[ht]
\centerline{
\includegraphics[width=0.85\linewidth]{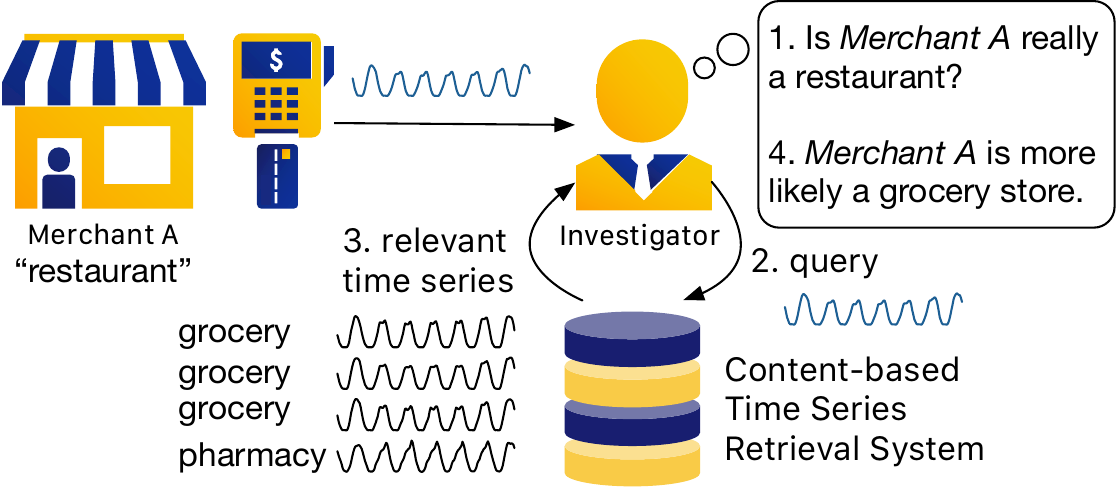}
}
\caption{
The use case for a CTSR system with transaction time series.
}
\label{fig:motivation_0}
\end{figure}

To evaluate the effectiveness and efficiency of different CTSR system designs in addressing the business problem presented in Figure~\ref{fig:motivation_0}, we constructed a transaction time series dataset for testing these CTSR solutions.

The dataset comprises 160,014 training time series, 19,993 test queries, and 19,992 validation queries, each representing the time series signature for a merchant, with a length of 168.
For a retrieved time series given a query time series, we consider the retrieved time series a relevant item from the database if it belongs to merchants of the same business type as the query time series.
If the retrieved time series is from another business type than the query time series, it is considered an irrelevant item.
We calculate the performance measurements at $k=10$ for each of the 19,993 test queries, and present the average results in Table~\ref{tab:visa_exact}. 
Only faster deep learning methods (i.e., LSTM, GRU, TF, RN1D, and RN2Dw/T) were tested on the transaction time series, as these methods were the clear winners from the experiments conducted on the UCR archive CTSR dataset, considering both effectiveness and efficiency.
The proposed RN2Dw/T method is the best performing method, with the difference in performance between it and the second-best RN1D method appearing small. 
However, the difference is statistically significant based on two-sample t-tests with $\alpha=0.05$.
All methods have similar query times (in seconds).

\begin{table}[ht]
\centering
\caption{
The proposed method outperforms the others in all performance measurements.
}
\label{tab:visa_exact}
\footnotesize
\begin{tabular}{l||ccc|c}
Method & PREC@10 & AP@10 & NDCG@10 & Query Time \\ \hline \hline
LSTM & 0.8798 & 0.8857 & 0.8830 & 0.0194 \\
GRU & 0.8503 & 0.8580 & 0.8545 & 0.0093 \\
TF & 0.8622 & 0.8678 & 0.8652 & 0.0040 \\
RN1D & 0.8941 & 0.8967 & 0.8955 & 0.0687 \\ \hline
RN2Dw/T & \textbf{0.8963} & \textbf{0.8999} & \textbf{0.8982} & 0.0197 \\
\end{tabular}
\end{table}





To provide a better user experience, we can further reduce the query time of the proposed RN2Dw/T system by replacing the exact nearest neighbor search algorithm with an approximate one. 
We compared the performance of the proposed system using both the exact and approximate nearest neighbor search algorithms.
For the approximate search, we utilized the nearest neighbor descent method as described in~\cite{dong2011efficient}, and implemented it using the PyNNDescent library~\cite{pynndescent}.
Our results indicate that the query time for the proposed method is 19.65 milliseconds for exact nearest neighbor search and 0.45 milliseconds for approximate nearest neighbor search. 
Moreover, the method's performances (i.e., PREC@10, AP@10, and NDCG@10) remain exactly the same as presented in Table~\ref{tab:visa_exact}.
In other words, we were able to improve the throughput by 43X without compromising the method's performance.

\section{Conclusion}
\label{sec:conclusion}
In this paper, we investigated the Content-based Time Series Retrieval (CTSR) problem and tested eight methods (ED, DTW, LSTM, GRU, TF, RN1D, RN2D, and RN2Dw/T).
Our results show that the proposed RN2Dw/T method outperformed the other methods in terms of both effectiveness and efficiency.
As part of our future work, we aim to enhance the efficiency of the proposed model by incorporating low-bit representation techniques, such as those presented in~\cite{wang2015learning,wang2017survey,yeh2022embedding,chen2022tinykg,andoni2022learning}.
Moreover, we have plans to expand the method's scope to accommodate multi-dimensional time series data. Additionally, we aim to tailor the method for application in an unsupervised setting with pre-training methods~\cite{yeh2023toward}.

\bibliographystyle{ACM-Reference-Format}
\bibliography{section/reference}

\end{document}